\documentclass[%
 aip,
 amsmath,amssymb,
 reprint,%
]{revtex4-1}

\usepackage{graphicx}
\usepackage{dcolumn}
\usepackage{bm}

\usepackage[utf8]{inputenc}
\usepackage[T1]{fontenc}
\usepackage{mathptmx}

\usepackage{multirow}

\begin{document}

\preprint{AIP/123-QED}

\title[SFs for rod-like particles]{Machine-learning effective many-body potentials for anisotropic particles using orientation-dependent symmetry functions}

\author{Gerardo Campos-Villalobos}
 \email{g.d.j.camposvillalobos@uu.nl}
\affiliation{ 
Soft Condensed Matter, Debye Institute for Nanomaterials Science, Utrecht University, Princetonplein 5, 3584 CC Utrecht, The Netherlands
}%
\author{Giuliana Giunta}%
\affiliation{ 
Soft Condensed Matter, Debye Institute for Nanomaterials Science, Utrecht University, Princetonplein 5, 3584 CC Utrecht, The Netherlands
}%

\author{Susana Mar\'in-Aguilar}%
\affiliation{ 
Soft Condensed Matter, Debye Institute for Nanomaterials Science, Utrecht University, Princetonplein 5, 3584 CC Utrecht, The Netherlands
}%

\author{Marjolein Dijkstra}%
\email{m.dijkstra@uu.nl}
\affiliation{ 
Soft Condensed Matter, Debye Institute for Nanomaterials Science, Utrecht University, Princetonplein 5, 3584 CC Utrecht, The Netherlands
}%

\date{\today}

\begin{abstract}
Spherically-symmetric atom-centered descriptors of atomic environments have been widely used for constructing potential or free energy surfaces of atomistic and colloidal systems and to characterize local structures using machine learning techniques. However, when particle shapes are non-spherical, as in the case of rods and ellipsoids, standard spherically-symmetric structure functions alone produce imprecise descriptions of local environments. In order to account for the effects of orientation, we introduce two- and three-body orientation-dependent particle-centered descriptors for systems composed of rod-like particles. To demonstrate the suitability of the proposed functions, we use an efficient feature selection scheme and simple linear regression to construct coarse-grained many-body interaction potentials for computationally-efficient simulations of model systems consisting of colloidal particles with anisotropic shape: mixtures of colloidal rods and nonadsorbing polymer, hard rods enclosed by an elastic microgel shell, and ligand-stabilized nanorods.  We validate the machine-learning (ML) effective many-body potentials based on orientation-dependent symmetry functions by using them in direct coexistence simulations to map out the phase behavior of colloidal rods and non-adsorbing polymer. We find good agreement with  results obtained from simulations of the true binary mixture, demonstrating that the  effective interactions are well-described by the orientation-dependent ML potentials.

\end{abstract}

\maketitle

\section{\label{sec:intro}Introduction}

Anisotropic molecules and colloids are able to self-assemble into complex structures with competing orientational and translational order, ranging from liquid crystals~\cite{care2005computer,allen2019molecular} to empty liquids.~\cite{ruzicka2011fresh}  Among colloidal systems, non-spherical particles have long been known and modern synthetic routes allow for a great variety of particles with different shape.~\cite{glotzer2007anisotropy} For example, in addition to rod-like particles of biological origin such as viruses,~\cite{wetter1985flussigkristalle,dogic1997smectic,grelet2014hard} colloidal rods can also be synthesized from a wide range of materials including  boehmite,~\cite{buining1994phase} $\beta$-ferric oxyhydroxide (FeOOH),~\cite{maeda2003liquid}  gold,~\cite{busbee2003improved} silica,~\cite{kuijk2012phase} cellulose,~\cite{schutz2015rod} and  titanium dioxide (TiO$_{2}$)  nanocrystals.~\cite{hosseini2020smectic} The assembly of such elongated particles has become increasingly popular because it allows for the formation of ordered superstructures exhibiting collective physical properties that depend on the shape and size of the constituent particles.~\cite{xie2011self} By precisely controlling physico-chemical parameters and boundary conditions of a self-assembly process, particle-based simulations can be especially effective in providing a clear insight into the link between the detailed interactions among particles/molecules and the resulting equilibrium properties.

Historically, there have been basically two approaches to model non-spherical particles: atomistic and coarse grained.~\cite{odriozola2011communication} In the former approach, chemistry-level detail is well represented at the expense of a larger computational cost. In the latter, many atoms are grouped together into single sites (beads, superatoms) which generally do not show a spherical symmetry. In such cases, generic single-site ellipsoidal pair potentials and simple hard-particle models have been widely used, thereby providing an understanding of the physics behind their mesoscopic and macroscopic behavior.~\cite{allen2019molecular} For an accurate representation of specific systems with complex interactions, developing predictive and computationally tractable coarse-grained (CG) models is necessary. However, this is not a trivial task. Difficulties arise in part because at such resolution level, the formal integration of a set of degrees of freedom leads to effective interaction potentials that typically require a description beyond the pairwise approximation.~\cite{bolhuis2001many,likos2001effective} The many-body terms in these CG potentials are in general very difficult to take into account,~\cite{likos2001effective} as it has been recognized in developing reduced-order models for star polymers,~\cite{watzlawek1999phase,von2000triplet} colloid-polymer mixtures~\cite{dijkstra2006effect,dijkstra1999phase} and ligand-stabilized nanoparticles.~\cite{schapotschnikow2009understanding,bauer2017three} In addition, even though the evaluation of these many-body potentials is in general computationally cheaper than simulating  the full or fine-grained system, the computational cost may still be high and  limit the range of accessible time- and length-scales.

In the past years, data-driven or machine learning (ML) approaches have been successfully applied to build accurate and computationally-efficient potentials for molecular and atomistic systems, and more generally to establish the relationship between a specific atomic configuration and the properties that can be computed by \emph{ab-initio} methods.~\cite{musil2021physics} In the field of soft matter and colloidal systems, a ML approach based on simple linear regression has been recently employed to efficiently represent many-body interactions in spherical microgel particles in 2D~\cite{boattini2020modeling} and to build an effective one-component interaction Hamiltonian for a mixture of colloidal hard spheres and non-adsorbing polymer.~\cite{campos2021machine}
In these approaches, the use of ML techniques was shown to serve as a powerful tool for speeding up by several orders of magnitude simulations that incorporate effective many-body interactions.

 \begin{figure*}
\includegraphics[scale=0.20]{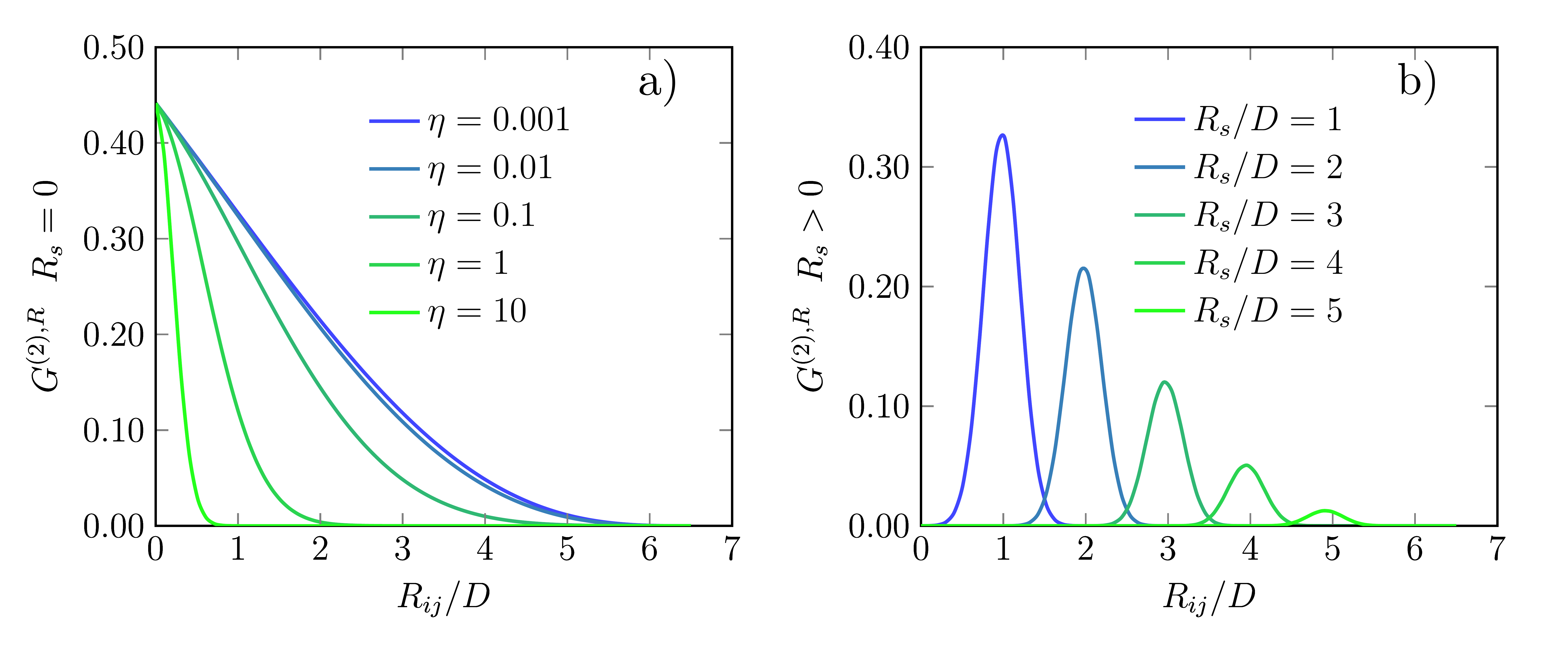}
\caption{\label{fig:GAR1} Two-body radial (R) symmetry function $G^{(2),R}$ as a function of the center-of-mass distance $R_{ij}/D$ between two (rod-like)  particles with diameter $D$. The  cut-off value is set to $r_{c}/D=6.5$. a) $G^{(2),R}$  for $R_{s}=0$ and different values of $\eta$ as labelled. b) $G^{(2),R}$ at fixed $\eta=10$ and for different shifting distances $R_{s}$ as labelled.}
\end{figure*}

The construction of ML potentials follows two main steps: the transformation of the atomic positions into suitable descriptors and the subsequent association of a (free) energy to this structure using a functional form provided by a ML method.~\cite{behler2011atom} In general terms, the \emph{descriptors} correspond to quantities that are easier to evaluate than the properties one ultimately aims at predicting, but that strongly correlate with the to-be-predicted properties.~\cite{musil2021physics} In practice, these descriptors are determined by applying geometric and algebraic operations on the Cartesian coordinates of the system, ultimately transforming them into mathematical objects that satisfy the conditions of smoothness and symmetry with respect to isometries.~\cite{musil2021physics} Typical representations based on descriptors of atomic/particle environments involve bond-order parameters and Fourier series of structural invariants.~\cite{harrington2019machine} However, the most commonly used representations include atom-centered symmetry functions  (SFs) ~\cite{behler2007generalized,behler2011atom} and the smooth overlap of atomic positions (SOAP).~\cite{bartok2013representing} For the purpose of constructing not only accurate but also computationally efficient potentials for anisotropic colloids or molecules, ML approaches such as those from Refs.~\onlinecite{boattini2020modeling,campos2021machine} are appealing. Nevertheless, available descriptors or structure functions such as the original atom-centered SFs by Behler and Parrinello~\cite{behler2007generalized} are, by construction, spherically-symmetric and do not take into account  orientation and alignment effects of non-spherical particles. Thus, in order to represent the many-body CG potentials in such systems using ML methods based on local structure characterization, suitable descriptors that capture this crucial aspect are needed. 

Within the recent efforts in correlating structure and dynamics in disordered systems, there have been some attempts in describing the local structure of systems composed of elongated particles by using spherically-symmetric SFs.~\cite{harrington2019machine} However, these approaches are simply based on describing  continuous elongated bodies such as ellipsoids by two distinct monomers. Hence, under such considerations, the individual structure functions for a reference elongated particle depend either on the position of the two individual monomers and the centroid of surrounding neighbors, or on a centroid and the two monomers composing a dimer of neighboring particles. Therefore, descriptors that are able to explicitly and simultaneously incorporate information on the orientation of a reference non-spherical particle and its neighbors, are still missing.

Here, we propose two- and three-body orientation-dependent particle-centered descriptors suitable for describing  local structure and  constructing many-body potentials for systems composed of rod-like particles, including spherocylinders, rigid linear chains, and ellipsoids.  Using a recently proposed feature selection scheme and  linear regression,~\cite{boattini2020modeling,campos2021machine} we construct effective many-body potentials for model systems of colloidal hard rods and non-adsorbing polymer and for core-shell microgel rods. Furthermore, the same approach is used to represent the effective orientation-dependent two-body potential of mean force (PMF) of ligand-stabilized nanorods.

The remainder of this article is organized as follows. In Section \ref{sec:model} we discuss the orientation-dependent descriptors for rod-like particles. The feature selection scheme and regression method used to construct the ML potentials for the three systems discussed above are briefly described in Section~\ref{sec:valida}. In this Section we also demonstrate the suitability of the particle-centered orientation-dependent descriptors for encoding information on the local structure of systems composed of prolate ellipsoidal particles. The accuracy of the coarse-grained ML potentials is further tested by directly comparing results obtained from Monte Carlo simulations of the "true" full systems with those coming from simulations using the reduced-order ML potentials. We conclude with a final discussion and remarks in Section \ref{sec:conclusions}.

\section{\label{sec:model}Symmetry functions for rod-like particles}

To introduce the orientation-dependent symmetry functions (ODSFs), we consider model systems of uniaxial particles with spherocylindrical shape (cylinders of length $L$ and diameter $D$ capped with two hemispheres at both ends) and ellipsoids of revolution for which two of the perpendicular semi-axes are equal, but different in general from the third: $\sigma_{||} \neq \sigma_{\perp} = \sigma_{\perp '}$. In these cases, each particle ${i}$ is characterized by its center-of-mass position vector, $\boldsymbol{R}_{i}$, and the orientation vector of its long axis, $\hat{\boldsymbol{u}}_{i}$. The descriptors discussed next are  a function of scalars that depend on these quantities. We note that, due to their general functional form, the ODSFs are also valid for rigid linear chains of arbitrary dimensions (fused- and tangent-sphere models) and for oblate ellipsoids with an infinite-fold rotational symmetry around the $\hat{\boldsymbol{u}}_{i}$ axis.

\begin{figure}
\includegraphics[scale=0.22]{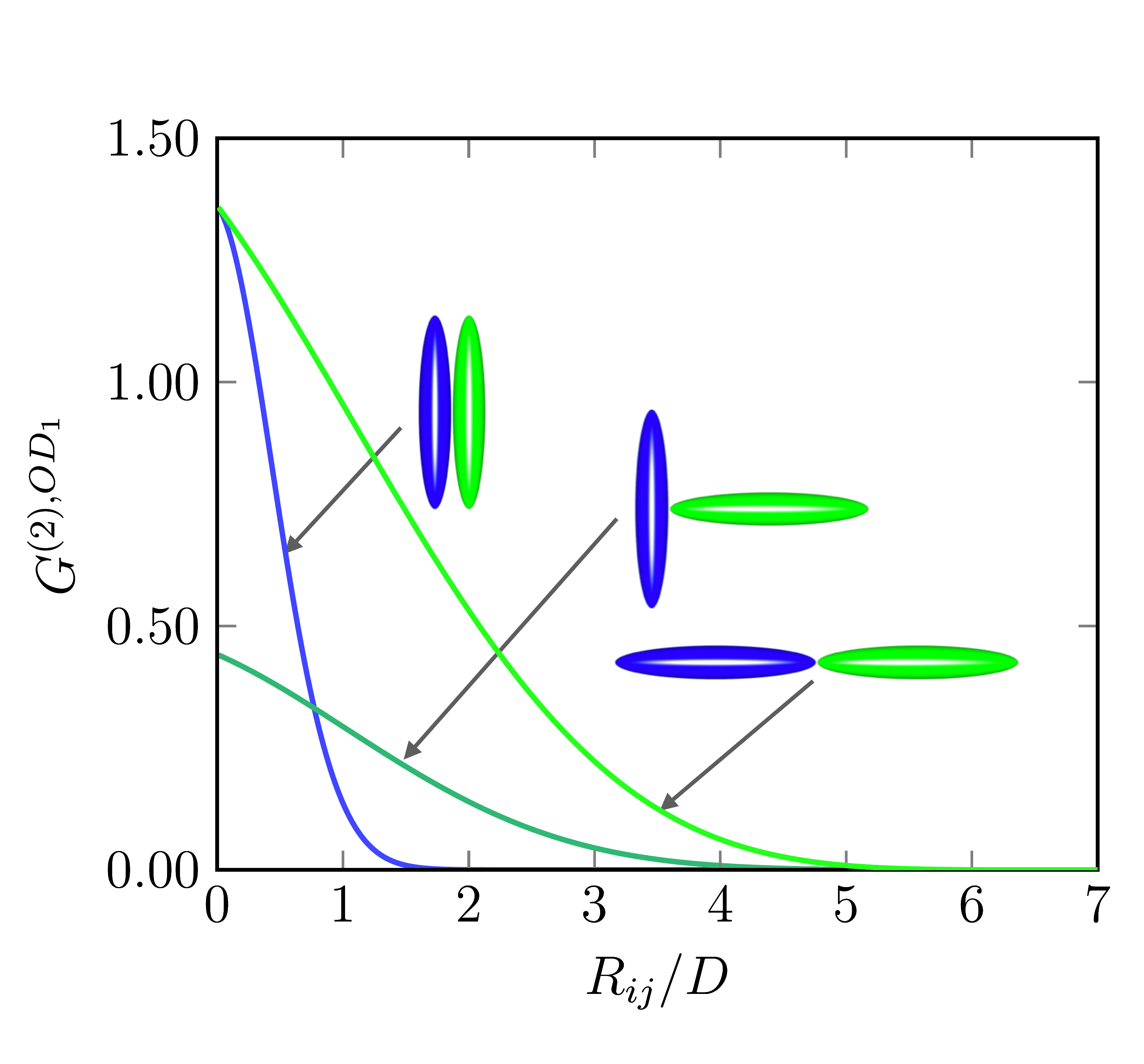}
\caption{\label{fig:GAR2} Two-body orientation-dependent  symmetry function $G^{(2),OD_1}$ as a function of the center-of-mass distance $R_{ij}/D$ between two rods with  diameter $D$. Three different relative orientations are considered. The function is shown for values $\sigma_{||}/D=3$ and $\sigma_{\perp}/D=0.5$. The cut-off value is set to $r_{c}/D=6.5$.}
\end{figure}

\begin{figure}
\includegraphics[scale=0.74]{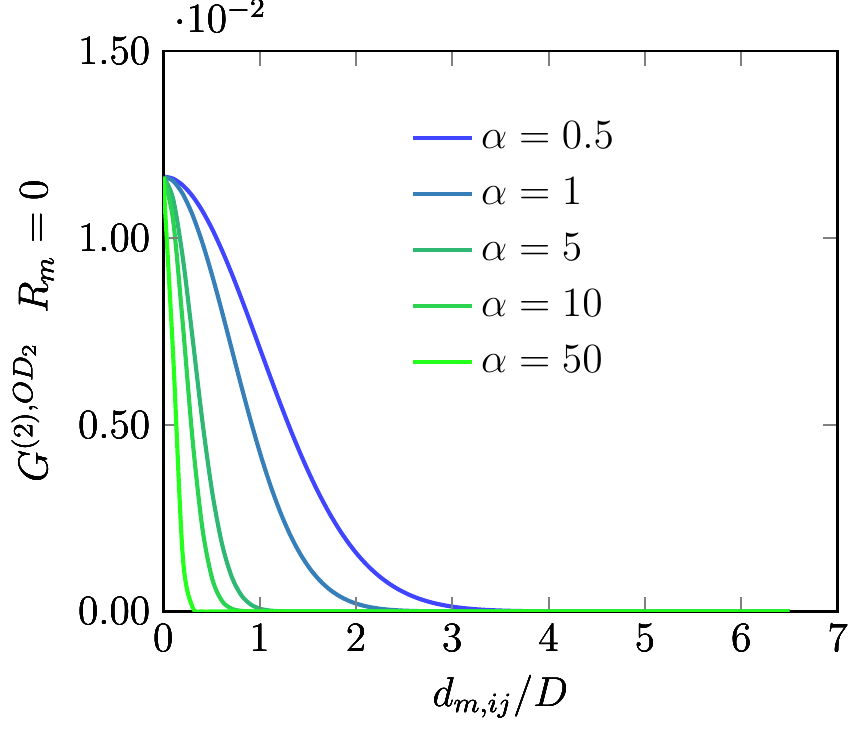}
\caption{\label{fig:GAR3} Two-body orientation-dependent   symmetry function $G^{(2),OD_2}$ as a function of the minimum distance $d_{m,ij}/D$ between the central axes of two rod-like particles with diameter $D$  for different values of $\alpha$ as labelled. The cut-off value is set to $r_{c}/D=6.5$ and the center-of-mass distance is fixed at $R_{ij}/D=5$.}
\end{figure}

We start the discussion of the functional form of the ODSFs by introducing a cutoff function $f_c(R_{ij})$: a monotonically decreasing function that smoothly goes to 0 in both value and slope at a cutoff distance $r_c$. Here, we consider a cutoff function of the form,
\begin{equation}
\label{cutf}
 f_{c}(R_{ij}) =
    \begin{cases}
      \tanh^{3} (1 - R_{ij}/r_{c}) & \text{for } R_{ij} \leq r_{c}\\
      0 & \text{for } R_{ij}>r_{c},
    \end{cases}       
\end{equation}
where $R_{ij}=|\boldsymbol{R}_{i}-\boldsymbol{R}_{j}|$ is the center-of-mass distance between particle $i$ and $j$ at positions $\boldsymbol{R}_{i}$ and $\boldsymbol{R}_{j}$, respectively. Each symmetry function discussed below is multiplied by one or more cutoff functions to ensure that the total symmetry function decays to zero in value and slope at the cutoff radius. Consequently, particles beyond the cutoff radius do not enter the reference particle contributions. We note that Eq.~\ref{cutf} was introduced in Ref.~\onlinecite{behler2011atom} and also used in Refs.~\onlinecite{singraber2019library,campos2021machine}.

\begin{figure*}[t]
\includegraphics[scale=0.20]{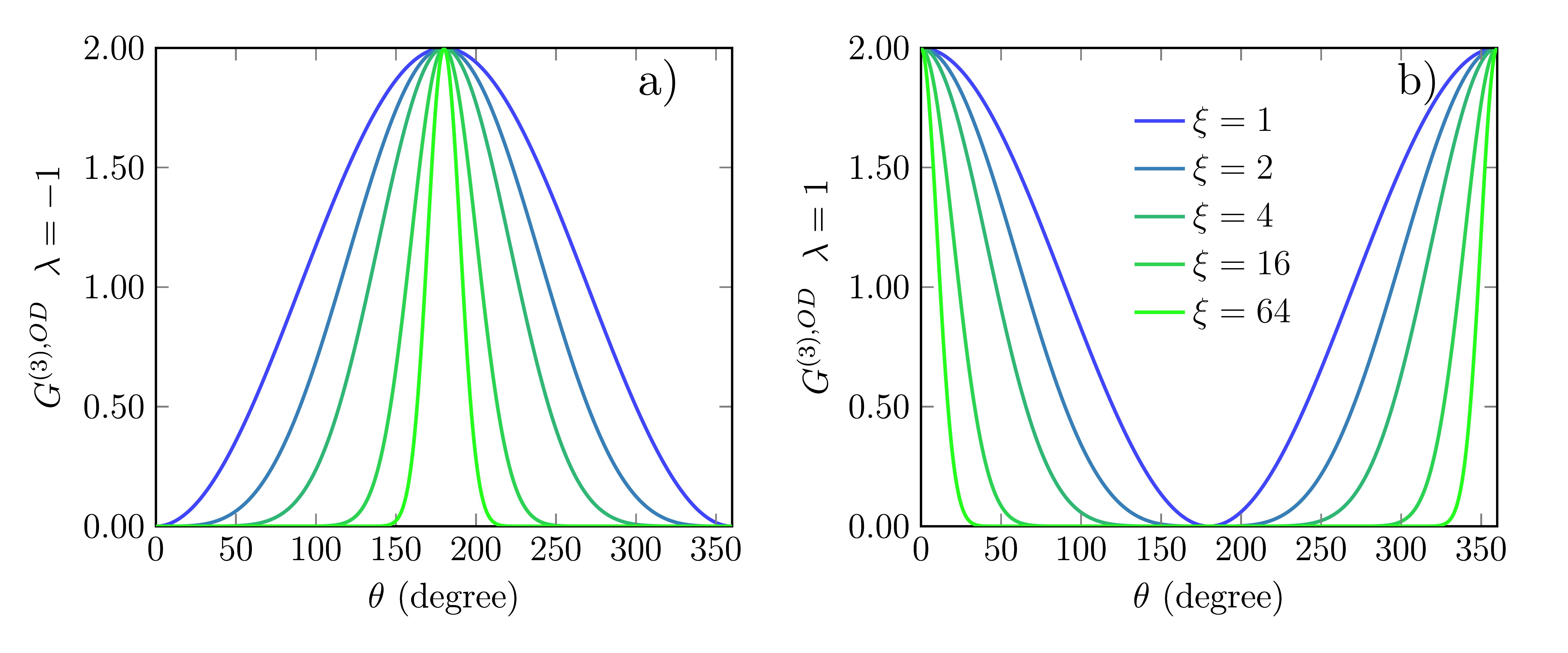}
\caption{\label{fig:GAA1} Angular contribution of the three-body orientation-dependent symmetry function $G^{(3),OD}$ for a particle with only two neighbors. a) $G^{(3),OD}$  for $\lambda=1$ and b) for $\lambda=-1$. The function is plotted for different values of $\xi$ as labelled.}
\end{figure*}

To describe the local environment of a rod-like particle we start from the spherically-symmetric two-body radial SFs $G^{(2),R}$ introduced by Behler and Parrinello for constructing high-dimensional neural network potentials,\cite{behler2007generalized}
\begin{equation}
    G^{(2),R}(i;\eta,R_{s})=\sum_{j}{e^{-\eta\left(R_{ij}-R_{s} \right)^{2}}\cdot f_{c}(R_{ij})},
\end{equation}
which is a sum of Gaussians multiplied by a cutoff function. The width of the Gaussians is defined by  parameter $\eta$, and the center of the Gaussian distributions can be shifted to a certain radial distance by the parameter $R_{s}$. In Fig.~\ref{fig:GAR1} we plot several example spherically-symmetric two-body radial SFs, $G^{(2),R}(i;\eta,R_{s})$,  for different sets of parameters. While these functions are able to capture two-body correlations between spherically-symmetric particles, they are not suitable for encoding information on the relative orientation of rod-like particles. To correctly describe the local environment of such anisotropic bodies, it is necessary to account for the orientational degrees of freedom. To this end, we introduce two orientation-dependent two-body SFs, $G^{(2),OD_1}$ and $G^{(2),OD_2}$, and one orientation-dependent  three-body one, $G^{(3),OD}$. The first ODSF is physically-motivated and is based on the assumption that for a reference particle $i$, the correlation with a neighboring particle $j$ occurs within an ellipsoid of revolution centered around particle $i$, with a variable spatial extent of the ellipsoid $2\sigma_{||}$ along the principal axis $\hat{\boldsymbol{u}}_{i}$ and $2\sigma_{\perp}$ perpendicular to it. By associating an anisotropic (trivariate) Gaussian to every ellipsoid, we thus compute a two-body function proportional to the mathematical overlap of the Gaussian of the reference particle and that of a neighbor. Such a function is efficiently represented by the  so-called "Overlap Model" potential as introduced by Berne and Pechukas.~\cite{berne1972gaussian} The resulting ODSF for particle $i$ is therefore a sum of cut-off functions multiplied by the overlap of its anisotropic Gaussian with those of neighboring particles $j$,

\begin{eqnarray}
    G^{(2),OD_1}(i;\sigma_{||},\sigma_{\perp}) &=& \nonumber \\ & & \hspace{-20mm} \sum_{j}{\epsilon(\hat{\boldsymbol{u}}_{i},\hat{\boldsymbol{u}}_{j})\cdot e^{-\left(R_{ij}/\sigma(\hat{\boldsymbol{u}}_{i},\hat{\boldsymbol{u}}_{j},\hat{\boldsymbol{R}}_{ij} ) \right)^{2}}\cdot f_{c}(R_{ij})},
\end{eqnarray}
where $\epsilon(\hat{\boldsymbol{u}}_{i},\hat{\boldsymbol{u}}_{j})$ and $\sigma(\hat{\boldsymbol{u}}_{i},\hat{\boldsymbol{u}}_{j},\hat{\boldsymbol{R}}_{ij} )$ are the  angle-dependent strength and range parameters, which for pairs of identical particles read
 \begin{equation}
 \label{eq:epsi}
     \epsilon(\hat{\boldsymbol{u}}_{i},\hat{\boldsymbol{u}}_{j}) = \left[ 1 - \chi^{2}(\hat{\boldsymbol{u}}_{i} \cdot \hat{\boldsymbol{u}}_{j})^{2}  \right]^{-1/2},
 \end{equation}

\begin{widetext}
\begin{equation}
\label{eq:sigma}
    \sigma(\hat{\boldsymbol{u}}_{i},\hat{\boldsymbol{u}}_{j},\hat{\boldsymbol{R}}_{ij} ) = \sigma \left(1 - \frac{1}{2} \chi \Bigg \{ \frac{\left(\hat{\boldsymbol{R}}_{ij}\cdot \hat{\boldsymbol{u}}_{i} + \hat{\boldsymbol{R}}_{ij}\cdot \hat{\boldsymbol{u}}_{j}   \right)^{2}}{\left[1 +\chi (\hat{\boldsymbol{u}}_{i} \cdot \hat{\boldsymbol{u}}_{j} ) \right]} +  \frac{\left(\hat{\boldsymbol{R}}_{ij}\cdot \hat{\boldsymbol{u}}_{i} - \hat{\boldsymbol{R}}_{ij}\cdot \hat{\boldsymbol{u}}_{j}   \right)^{2}}{\left[1 -\chi (\hat{\boldsymbol{u}}_{i} \cdot \hat{\boldsymbol{u}}_{j} ) \right]}\Bigg\} \right)^{-1/2},
\end{equation}
\end{widetext}
where $\hat{\boldsymbol{R}}_{ij}=\boldsymbol{R}_{ij}/R_{ij}$ is the unit vector along the center-of-mass distance  vector between particle $i$ and $j$, $\boldsymbol{R}_{ij} = \boldsymbol{R}_{i}-\boldsymbol{R}_{j}$, $\sigma=\sqrt{2} \sigma_{\perp}$ and $\chi=(\sigma_{||}^{2} - \sigma_{\perp}^{2})/(\sigma_{||}^{2} + \sigma_{\perp}^{2})$ are the range and anisotropy parameters, respectively. The two parameters that control the shape of this function are the spatial extents of the ellipsoid of revolution along the principal axis and perpendicular to it, i.e.  $\sigma_{||}$ and $\sigma_{\perp}$. This function carries a direct dependence on the relative orientation of the two rods  and is sensitive enough to simultaneously capture information on the inter-particle distances as can be appreciated from Fig.~\ref{fig:GAR2}.

The second ODSF considered here, corresponds to a generalization of the $G^{(2),R}(i;\eta,R_{s})$ function to  non-spherical particles, where $R_{ij}$ is replaced by the minimum distance between the long axes of the two particles, $d_{m,ij}(\boldsymbol{R}_{ij},\hat{\boldsymbol{u}}_{i},\hat{\boldsymbol{u}}_{j})$.~\footnote{A fast algorithm to calculate $d_{m,ij}$  for two bodies of spherocylindrical symmetry  has been described by Vega and Lago.~\cite{vega1994fast}} As in the case of the $G^{(2),R}(i;\eta,R_{s})$-type functions, the width of the Gaussians is defined by  parameter $\alpha$, and the center of the Gaussian distributions can be shifted by parameter $R_{m}$
\begin{equation}
    G^{(2),OD_2}(i;\alpha, R_{m}) = \sum_{j}{e^{-\alpha\left(d_{m,ij}(\boldsymbol{R}_{ij},\hat{\boldsymbol{u}}_{i},\hat{\boldsymbol{u}}_{j})-R_{m} \right)^{2}}\cdot f_{c}(R_{ij})}.
\end{equation}

In Fig.~\ref{fig:GAR3}, we exemplify the plot of  $G^{(2),OD_2}(i;\alpha, R_{m})$ for varying parameters.

Finally, we also introduce an angular three-body ODSF $G^{(3),OD}(i;\sigma_{||}, \sigma_{\perp}, \lambda, \xi)$, that depends on the angle $\theta_{ijk} = \arccos(\boldsymbol{R}_{ij} \cdot \boldsymbol{R}_{ik}/(R_{ij} R_{ik}))$ centered at reference particle $i$,
\begin{widetext}
\begin{eqnarray}
    G^{(3),OD}(i;\sigma_{||}, \sigma_{\perp}, \lambda, \xi) =&& 2^{1-\xi}\sum_{j,k\neq i}\left( 1 + \lambda \cos \theta_{ijk} \right)^\xi \epsilon_{ij}(\hat{\boldsymbol{u}}_{i},\hat{\boldsymbol{u}}_{j}) \epsilon_{ik}(\hat{\boldsymbol{u}}_{i},\hat{\boldsymbol{u}}_{k})
    \epsilon_{jk}(\hat{\boldsymbol{u}}_{j},\hat{\boldsymbol{u}}_{k}) \nonumber\\
    && \times e^{-\left(R_{ij}^{2}/\sigma_{ij}^{2}(\hat{\boldsymbol{u}}_{i},\hat{\boldsymbol{u}}_{j},\hat{\boldsymbol{R}}_{ij}) +R_{ik}^{2}/\sigma_{ik}^{2}(\hat{\boldsymbol{u}}_{i},\hat{\boldsymbol{u}}_{k},\hat{\boldsymbol{R}}_{ik})+ R_{jk}^{2}/\sigma_{jk}^{2}(\hat{\boldsymbol{u}}_{j},\hat{\boldsymbol{u}}_{k},\hat{\boldsymbol{R}}_{jk})\right)} \nonumber\\ &&\times f_{c}(R_{ij})f_{c}(R_{ik})f_{c}(R_{jk}),
\label{eq:g3}
\end{eqnarray}
\end{widetext}
where the indices $j$ and $k$ run over all the neighbors of particle $i$, and $\xi$ and $\lambda$ are two parameters that determine the shape of the function. The parameter $\lambda$ can have the values $+1$ or $-1$ and determines the angle $\theta_{ijk}$ at which the angular part of the function has its maximum. The angular resolution is provided by the parameter $\xi$, while the terms $\sigma_{ab}(\hat{\boldsymbol{u}}_{a},\hat{\boldsymbol{u}}_{b},\hat{\boldsymbol{R}}_{ab})$ and $\epsilon_{ab}(\hat{\boldsymbol{u}}_{a},\hat{\boldsymbol{u}}_{b})$ control the radial resolution via the anisotropy parameters as defined above. The angular part of this function for $\lambda=1$ and $-1$ and different $\xi$ values is shown in Fig.~\ref{fig:GAA1}.

The ODSFs introduced here provide a rotationally and translationally invariant description of the environment because they depend on the internal coordinates $R_{ij}$, $d_{m,ij}$ $\theta_{ijk}$ and scalar products of pairs of vectors. Because of the sum over all neighbors within $r_{c}$, they are invariant with respect to any permutation of equivalent particles in the environment. As it will become apparent in Section~\ref{sec:valida}, when describing the local environment of elongated particles, several ODSFs which carry different structural information are combined together. Therefore, the unique significance of the individual functions is generally lost, but when they are used as a group a more complete and unbiased description of the local structure is achieved.

\section{\label{sec:valida}Machine-learning potentials for rod-like particles}

In this section, we demonstrate the suitability of the ODSFs for rod-like particles as descriptors for constructing ML coarse-grained potentials for varying systems with different levels of complexity. More specifically, in order to show the generality of the approach, we chose model systems of prolate Gay-Berne particles, mixtures of spherocylindrical colloids and non-adsorbing polymer, core-shell microgel rods and ligand-stabilized nanorods. By selecting these models, we cover anisotropic particles represented as prolate ellipsoids, spherocylinders and rigid linear chains of fused and tangent sites. In all cases, we construct ML potentials as linear combinations of a number of symmetry functions $N_{SF}$. Here, the goal is to express the total effective potential of a system of $N$ non-spherical particles with positions $\{\boldsymbol{R}_{i}\}$ and orientations $\{\hat{\boldsymbol{u}}_{i}\}$ as 

\begin{equation}
    \Phi(\{\boldsymbol{R}_{i}, \hat{\boldsymbol{u}}_{i} \}) = \sum_{i}^{N}\sum_{k}^{N_{SF}} \omega_{k} G_{k}(i), 
\end{equation}
where $G_{k}(i)$ is the $k$-th descriptor describing the local environment of particle $i$ and $\omega_{k}$ the coefficient (weight) of the corresponding SF, which is fixed by the fitting procedure. Note that since $\Phi(\{\boldsymbol{R}_{i}, \hat{\boldsymbol{u}}_{i} \})$ is expressed as a sum of per-particle contributions, the fitting of an $N$-particle system can be extended to simulations with a different number of particles. We select the optimal subset of descriptors using the feature selection scheme of Ref.~\onlinecite{boattini2020modeling}, which we briefly summarize below.

For a given data set at hand, which typically consists of a collection of particle configurations and the corresponding values of the to-be-predicted quantities (e.g. energy), a training/validation split of the whole data is used. We note that unless stated otherwise, a 80/20 training/validation split of the whole data sets is adopted here. The first step of the method involves the creation of a large but manageable pool of candidate SFs. This is done by calculating, for every particle in the different configurations in the training data set, $M$ SFs with different sets of parameters. Then, an optimal subset of $N_{SF}<M$ SFs is selected from the pool in a step-wise fashion. The first SF that is selected corresponds to the one with the largest correlation with the target function as quantified by the square of the Pearson correlation coefficient, defined as

\begin{equation}
    c_{k}=\frac{ \sum_{j}{\left (\sum_i G_k(i)|_{j} -\overline{  \sum_i G_k(i)}\right)\left(\phi|_{j}-\overline{ \phi}\right)}}{\sigma_{\text{SD}}(\sum_i G_k(i)) \sigma_{\text{SD}}(\phi)},
\end{equation}
where $\sum_i G_k(i)|_{j}$ represents the sum of the $k$-th SF over all particles $i$ in configuration $j$ and $\phi|_{j}$ denotes the target variable evaluated for this configuration. $\overline{  \sum_i G_k(i)}$ and $\overline{ \phi}$ correspond to arithmetic means over all the configurations in the data set, and $\sigma_{\text{SD}}(\sum_i G_k(i))$ and $\sigma_{\text{SD}}(\phi)$ to their standard deviations. The next SF is then selected based on the highest increase  in the linear correlation between the currently selected set and the target data as determined by the coefficient of multiple correlation

\begin{equation} 
R^2 = {\bf c}^T {\bf R}^{-1} {\bf c},
\end{equation}
where ${\bf c}^T = (c_1, c_2,\cdots)$ is the vector whose $i$-th component is given by the Pearson correlation coefficient, $c_{i}$, between the $i$-th SF and the target data, and ${\bf R}$ is the correlation matrix of the current set of SFs with elements $\mathcal{R}_{ij}$ representing the Pearson correlation function between the $i$-th and $j$-th SF. In the case of only one SF, $R^{2}$ reduces to $c_{i}^{2}$. We note that $R^{2}$ can also be computed as the fraction of variance that is explained by a linear fit (including an intercept) of the target function in terms of the SFs in the set. The latter way of computing $R^{2}$ turns out to be slightly more expensive, but has the advantage of being numerically more stable.~\cite{boattini2020modeling} 

Maximizing the increase in the linear correlation with the target variable guarantees that only SFs that add relevant information are selected.~\cite{boattini2020modeling} This process is repeated iteratively and new SFs are selected until the correlation stops increasing appreciably.  This, indicates that the remaining SFs in the pool add negligible (irrelevant) information to the model. In turn, this constitutes a simple rule to optimize the number of selected SFs as their inclusion  would simply imply an unnecessary numerical overhead. All the parameters employed to generate the pool of SFs that are used to build the CG effective potentials for the different systems discussed next are reported in the Supplementary Material. We note that (if sufficient) using linear regression instead of other more complex schemes such as nonlinear neural networks, might have some important advantages, namely, the deterministic against the stochastic optimization of model parameters, the control on the number of features and the final reduced computational cost.~\cite{boattini2020modeling} The latter aspect, as it can be inferred, is in part determined by the optimal value of $N_{SF}$ and the type of descriptor used in the ML potential. In the Supplementary Material, we briefly discuss the computational cost of the potentials constructed for the systems introduced next.

\subsection{\label{sec:GB} Application to ellipsoids: The Gay-Berne model}

To test the performance of the orientation-dependent descriptors we start by constructing interaction potentials of particles with ellipsoidal symmetry. In particular, for instructive purposes, we use the well-known Gay-Berne (GB) model, in which the anisotropic pair potential reads,
\begin{equation}
    \phi_{\text{GB}}(\boldsymbol{R}_{ij}, \hat{\boldsymbol{u}}_{i},\hat{\boldsymbol{u}}_{j}) = 4 \varepsilon'(\hat{\boldsymbol{R}}_{ij}, \hat{\boldsymbol{u}}_{i},\hat{\boldsymbol{u}}_{j}) \left[ \rho_{ij}^{-12} - \rho_{ij}^{-6} \right],
\end{equation}
where 
\begin{equation}
 \rho_{ij} = \frac{R_{ij} - \sigma(\hat{\boldsymbol{u}}_{i},\hat{\boldsymbol{u}}_{j},\hat{\boldsymbol{R}}_{ij}) + \sigma_{0} }{\sigma_{0}},
\end{equation}
$R_{ij}$ is the distance between the centres of mass of particle $i$ and $j$, and $\hat{\boldsymbol{R}}_{ij}=\boldsymbol{R}_{ij}/R_{ij}$ is the unit vector along the separation vector $\boldsymbol{R}_{ij} = \boldsymbol{R}_{i}-\boldsymbol{R}_{j}$. The anisotropic contact distance $\sigma(\hat{\boldsymbol{u}}_{i},\hat{\boldsymbol{u}}_{j},\hat{\boldsymbol{R}}_{ij})$ and depth of the interaction energy $\varepsilon'(\hat{\boldsymbol{R}}_{ij}, \hat{\boldsymbol{u}}_{i},\hat{\boldsymbol{u}}_{j})$ depend on the orientational unit vector, length-to-breadth ratio ($\kappa=\sigma_{||}/\sigma_{\perp}$) and the energy depth anisotropy ($\kappa'=\epsilon_{\perp}/\epsilon_{||}$), which correspond to the ratio of the size and energy parameters in the end-to-end ($||$) and side-by-side ($\perp$) configurations. The contact distance function is given by Eq.~\ref{eq:sigma}, while the depth interaction energy reads
\begin{equation}
    \varepsilon'(\hat{\boldsymbol{R}}_{ij}, \hat{\boldsymbol{u}}_{i},\hat{\boldsymbol{u}}_{j}) = \epsilon \times \left[ \epsilon(\hat{\boldsymbol{u}}_{i},\hat{\boldsymbol{u}}_{j} )\right]^{\nu} \times \left[ \epsilon_{1}(\hat{\boldsymbol{R}}_{ij},\hat{\boldsymbol{u}}_{i},\hat{\boldsymbol{u}}_{j}) \right]^{\mu},
\end{equation}
where $\epsilon(\hat{\boldsymbol{u}}_{i},\hat{\boldsymbol{u}}_{j} )$ is defined by Eq.~\ref{eq:epsi},
\begin{widetext}
\begin{equation}
 \epsilon_{1}(\hat{\boldsymbol{R}}_{ij},\hat{\boldsymbol{u}}_{i},\hat{\boldsymbol{u}}_{j}) = 1 -   \frac{\chi'}{2} \left[ \frac{(\hat{\boldsymbol{R}}_{ij} \cdot \hat{\boldsymbol{u}}_{i} + \hat{\boldsymbol{R}}_{ij} \cdot \hat{\boldsymbol{u}}_{j})^{2}}{\left[ 1 + \chi' (\hat{\boldsymbol{u}}_{i} \cdot \hat{\boldsymbol{u}}_{j})\right]} + \frac{(\hat{\boldsymbol{R}}_{ij} \cdot \hat{\boldsymbol{u}}_{i} - \hat{\boldsymbol{R}}_{ij} \cdot \hat{\boldsymbol{u}}_{j})^{2}}{\left[ 1 - \chi' (\hat{\boldsymbol{u}}_{i} \cdot \hat{\boldsymbol{u}}_{j})\right]} \right],
\end{equation}
\end{widetext}
and $\chi'=\left[(\kappa')^{1/\mu} - 1 \right]/ \left[ (\kappa')^{1/\mu} + 1\right]$. In the equations above, $\varepsilon$ and $\sigma_{0}$ represent, respectively, the energy and length scales of the interaction. Here, we use the well-known GB model, with characteristic parameters ($\kappa, \kappa', \mu, \nu$)=($3.0,5.0,2.0,1.0$), which were originally used by Gay and Berne.~\cite{gay1981modification}

\begin{figure*}
\includegraphics[scale=0.205]{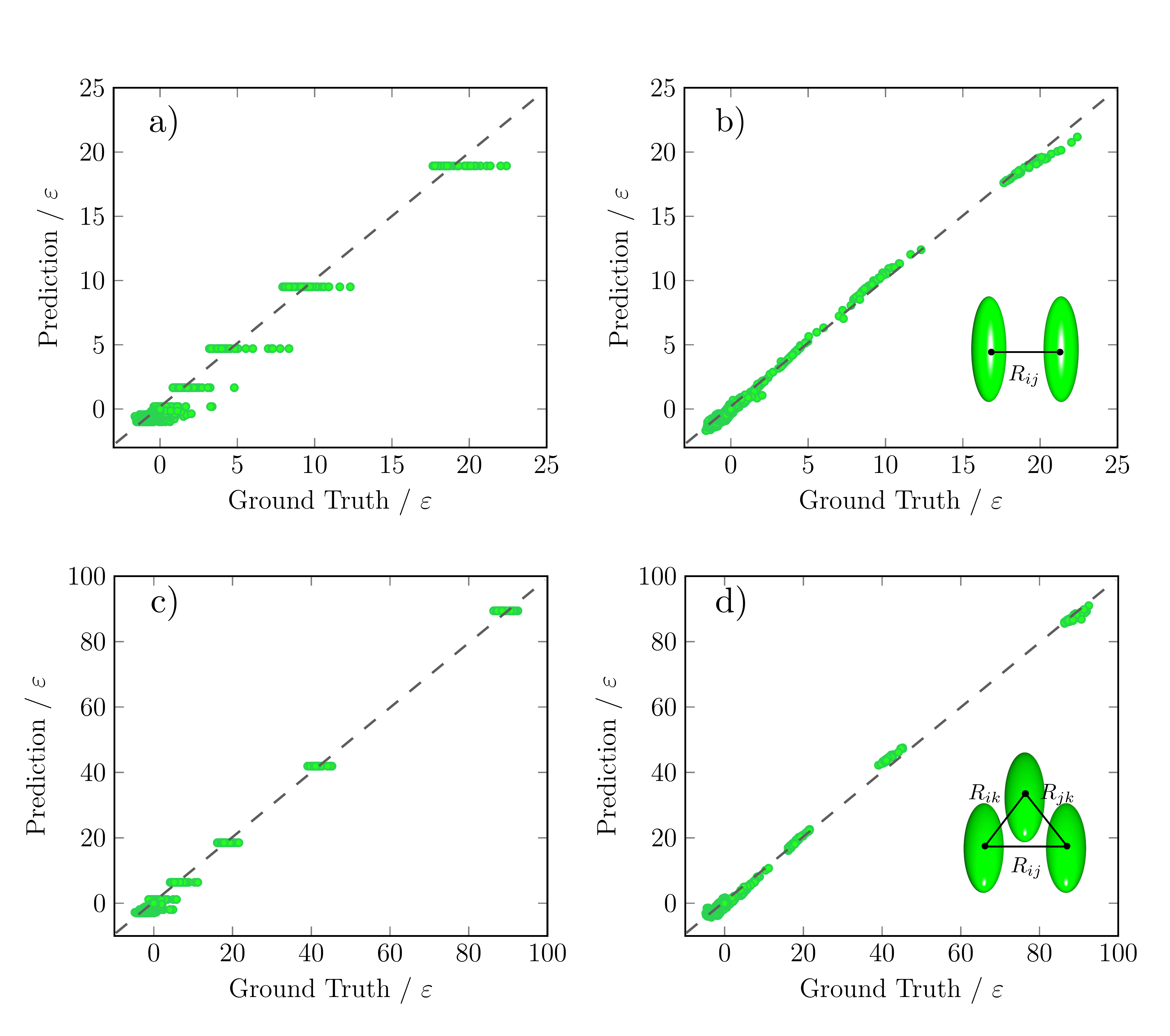}
\caption{\label{fig:GBD} Parity plots comparing total interaction energies of the true Gay-Berne model with characteristic parameters ($\kappa, \kappa', \mu, \nu$) = ($3.0,5.0,2.0,1.0$) and the ML model constructed using only two-body radial $G^{(2),R}$ SFs (left panels) and  with two-body orientation-dependent $G^{(2),OD_1}$ SFs (right panels). In both models, the number of descriptors is $N_{SF}=70$. a) and b) show results  for the dimer configurations, and c) and d)  report the comparison for the trimer configurations.}
\end{figure*}

In particular, we focus on fitting the potential energies $\phi$ of both dimer and trimer configurations (confined to equilateral triangles only, i.e. $R_{ij}=R_{ik}=R_{jk}$). To generate a training data set, we perform Monte Carlo (MC) simulations of  dimers and trimers at reduced temperature $T^{*}\equiv k_{\text{B}}T/\varepsilon=1.0$, where we fix the particle positions, but allow them to freely rotate in space and measure their total potential energy in 250 different samples collected from runs of $2.5\times10^{6}$ MC cycles. For each system, 100 separation distances in the range $0.85 \leq R_{ij}/\sigma_{0} \leq 4.95$ are considered. The selected range of separation distances contains both high-energy (repulsive) and low-energy (attractive) configurations. With this choice, each set contains a total of 25,000 samples, from which $80\%$ are used for training and $20\%$ are reserved for validation. We perform two different fits: (i) using two-body radial $G^{(2), R}$ SFs, which only depend on the distance between centroids of particles and (ii) using two-body orientation-dependent $G^{(2), OD_1}$ ODSFs. In both cases we restrict the number of descriptors to $N_{SF}=70$ and set the cut-off value at $r_c/\sigma_{0}=6.5$. Parity plots, comparing the original ground truth (training and validation) and predicted potential energies using the two ML models for the dimers and trimers are reported in Fig.~\ref{fig:GBD}. Since our data sets contain configurations at fixed $R_{ij}$ and varying particle orientations (for which the potential energy differs slightly), small clouds of points are identified. The models constructed with only $G^{(2), R}$ SFs show correlation coefficients of $R^{2} \approx 0.9$  and root mean square errors (RMSE), which are defined by the simple relation $R^2=1 - \text{RMSE}^2/\sigma^2_{SD}(\phi)$, of approximately $0.30\epsilon$ and $0.55\epsilon$ on both the training and validation sets of trimers and dimers, respectively. However, as it can be expected, they present a critical defect as shown in Fig.~\ref{fig:GBD}. In particular, since the descriptors are unable to encode information on the orientation-dependence of the interactions, the models assign the same energy to configurations with identical $R_{ij}$ but with different relative orientation of the particles. In contrast, when $G^{(2), OD_1}$ SFs are employed, the resulting models not only present correlation coefficients of $R^{2} \approx 0.99$ and RMSE values of $0.11\epsilon$ and $0.46\epsilon$ for trimers and dimers, respectively, but they able to capture accurately the orientation-dependence of the interaction energy. If the dimer and trimer data sets are combined and fitted simultaneously, the quality of the resulting model with the same number of features remains virtually unaffected.

\subsection{\label{sec:mix} Effective one-component Hamiltonian for colloidal hard rods and nonadsorbing polymer}

We continue our discussion by applying the ODSFs for constructing a direct relationship between the structure and effective many-body interactions in systems of rod-like particles with spherocylindrical shape. Such a particle shape has been widely used to represent generic colloidal nanorods and to investigate their phase behavior and self-assembling behavior.~\cite{bolhuis1997tracing,martinez2001gibbs,campos2021nonconventional} Here, we consider a system of sterically stabilized colloidal rods and non-adsorbing polymer. By departing from the thermodynamic potential of the full binary mixture, we formally reduce the problem to a colloids-only effective Hamiltonian, which incorporates many-body effects. A ML model is then constructed to represent such an effective Hamiltonian as a function of all colloid coordinates and orientations.
 
A simple model for the mixture is the so-called Asakura-Oosawa (AO) model, where the colloids are treated as hard particles, while the non-interacting polymer coils are regarded as point particles, which are excluded from the surface of the colloids by a distance equal to the radius of gyration of the polymer $R_{g}$.~\cite{asakura1958interaction,vrij1976polymers,dijkstra1999phase,dijkstra2006effect} The diameter of the coils is $\sigma_{p}=2R_{g}$. The colloids are represented by hard spherocylinders, which consist of cylinders of diameter $\sigma_{c}$ and length $L$ with semi-spherical caps at both ends with diameter $\sigma_{c}$. The pair potentials of this model are given by
\begin{equation}
\phi_{cc}(\boldsymbol{R}_{ij},\hat{\boldsymbol{u}}_{i},\hat{\boldsymbol{u}}_{j}) =
    \begin{cases}
      \infty & \text{for } d_{m,ij}(\boldsymbol{R}_{ij},\hat{\boldsymbol{u}}_{i},\hat{\boldsymbol{u}}_{j}) < \sigma_{c}\\
      0 & \text{otherwise,}
    \end{cases}       
\end{equation}

\begin{equation}
\phi_{cp}(\boldsymbol{R}_{i}-\boldsymbol{r}_{j},\hat{\boldsymbol{u}}_{i}) =
    \begin{cases}
      \infty & \text{for } d_{m,ij}(\boldsymbol{R}_{i}-\boldsymbol{r}_{j},\hat{\boldsymbol{u}}_{i}) < \sigma_{cp}\\
      0 & \text{otherwise,}
    \end{cases}       
\end{equation}

\begin{equation}
    \phi_{pp}(r_{ij})=0,
\end{equation}
where $\sigma_{cp}=(\sigma_{c}+\sigma_{p})/2$, $\boldsymbol{R}_{ij}=\boldsymbol{R}_{i}-\boldsymbol{R}_{j}$ with $\boldsymbol{R}_{i}$ and $\boldsymbol{R}_{j}$ the center-of-masses of  spherocylinder $i$ and $j$, respectively, and  $d_{m,ij}(\boldsymbol{R}_{ij},\hat{\boldsymbol{u}}_{i},\hat{\boldsymbol{u}}_{j})$ the minimum distance between the central axes of the two spherocylinders with orientations $\hat{\boldsymbol{u}}_{i}$ and $\hat{\boldsymbol{u}}_{j}$, $d_{m,ij}(\boldsymbol{R}_{i}-\boldsymbol{r}_{j},\hat{\boldsymbol{u}}_{i})$ the minimum distance between the spherocylinder axis and the polymer center-of-mass at $\boldsymbol{r}_{j}$ in the case of colloid-polymer interactions, and $r_{ij}=| \boldsymbol{r}_{i}- \boldsymbol{r}_{j} |$ the distance between the two polymer center-of-masses.

The total interaction Hamiltonian of the mixture of $N_{c}$ colloidal rods and $N_{p}$ polymer coils in a volume $V$ at absolute temperature $T$ reads $H=H_{cc}+H_{cp}+H_{pp}$, where $H_{cc}=\sum_{i<j}^{N_{c}}\phi_{cc}(\boldsymbol{R}_{ij},\hat{\boldsymbol{u}}_{i},\hat{\boldsymbol{u}}_{j})$, $H_{cp}=\sum_{i=1}^{N_{c}}\sum_{j=1}^{N_{p}}\phi_{cp}(\boldsymbol{R}_{i}-\boldsymbol{r}_{j},\hat{\boldsymbol{u}}_{i})$ and $H_{pp}=\sum_{i<j}^{N_{p}}\phi_{pp}(r_{ij})=0$. By keeping constant the number of colloids $N_{c}$ and treating the polymer coils grand-canonically, in which the polymer fugacity $z_{p}$ is fixed, the thermodynamic potential of the binary mixture, $F(N_{c},z_{p},V,T)$, can be written as a function of an effective one-component (only colloids) Hamiltonian where the polymer coils are formally integrated out
\begin{equation}
\label{f}
\exp\lbrack -\beta F \rbrack=  \frac{1}{N_{c}!\Lambda_{\text{c}}^{3N_{c}}}\int_V  d\boldsymbol{R}^{N_{c}} \int d\hat{\boldsymbol{u}}^{N_{c}}  \exp\lbrack -\beta H_{\text{eff}}\rbrack, 
\end{equation}
where $\beta = (k_{\text{B}}T)^{-1}$ with $k_{\text{B}}$ the Boltzmann constant and $H_{\text{eff}}= H_{cc}+ \Omega$, with $\Omega$ the grand potential of a ``sea" of ideal polymer at fugacity $z_{p}$ in the external field of a fixed configuration of $N_{c}$ colloidal rods.~\cite{dijkstra1999phase,dijkstra2006effect,dijkstra2002entropic} For the AO model, the grand potential $\Omega$ reads 
\begin{equation}
\label{om_eq}
\Omega = -z_{p} V_{f}( \{\boldsymbol{R}_i, \hat{\boldsymbol{u}}_{i}\}),
\end{equation}
where $V_{f}( \{\boldsymbol{R}_i, \hat{\boldsymbol{u}}_{i}\})$ is the free volume available for the polymer in the static configuration of $N_{c}$ colloidal rods with positions $\{ \boldsymbol{R}_i \}$ and orientations $\{ \hat{\boldsymbol{u}}_{i} \}$ and with  $i=1,...,N_{c}$. $V_{f}$ can be seen as the volume outside the $N_{c}$ depletion zones and can be decomposed into a zero-, one-, two-, three-, and higher-body contribution, $ V_{f}= V_{f}^{(0)} + \sum_{i=1}^{N_{c}}{V_{f}^{(1)}(\boldsymbol{R}_{i},\hat{\boldsymbol{u}}_{i})} + \sum_{i<j}^{N_{c}}{V_{f}^{(2)}(\boldsymbol{R}_{i}, \boldsymbol{R}_{j},\hat{\boldsymbol{u}}_{i},\hat{\boldsymbol{u}}_{j})} + V_{f}^{(3+)}$. In turn, for a fixed colloid configuration, the volume outside the $N_{c}$ depletion zones will be determined by the size ratio  $q=\sigma_{p}/\sigma_{c}$ between the polymer coils and colloids. For $q < 0.1547$, a  mapping onto an effective one-component system with an effective Hamiltonian based on pairwise additive depletion potentials is exact.~\cite{gast1983polymer,dijkstra1999phase,savenko2006phase} However, for larger $q$, the many-body contributions in  $V_{f}^{(3+)}$ must be considered.

For spherical colloids, the effective pair potential is given by the AO depletion potential which is known analytically.~\cite{dijkstra2006effect} However, no analogous analytic expression exists for the overlap volume of two
depletion zones of finite spherocylinders with arbitrary orientations and positions and, hence,  $V_{f}^{(2)}$ can only be approximated as in Ref.~\onlinecite{savenko2006phase} or calculated numerically. It turns out that the numerical evaluation of $V_{f}^{(2)}$ is almost as expensive as that of the whole free volume $V_{f}$, which prevents us from using such an approach in long simulations of large systems. Therefore, in order to construct for the first time a full many-body effective one-component interaction Hamiltonian for such a  mixture which is computationally-efficient, we fit $V_{f}$ as a function of all colloid coordinates using a set of ODSFs as discussed above.

\begin{figure}
\includegraphics[scale=0.74]{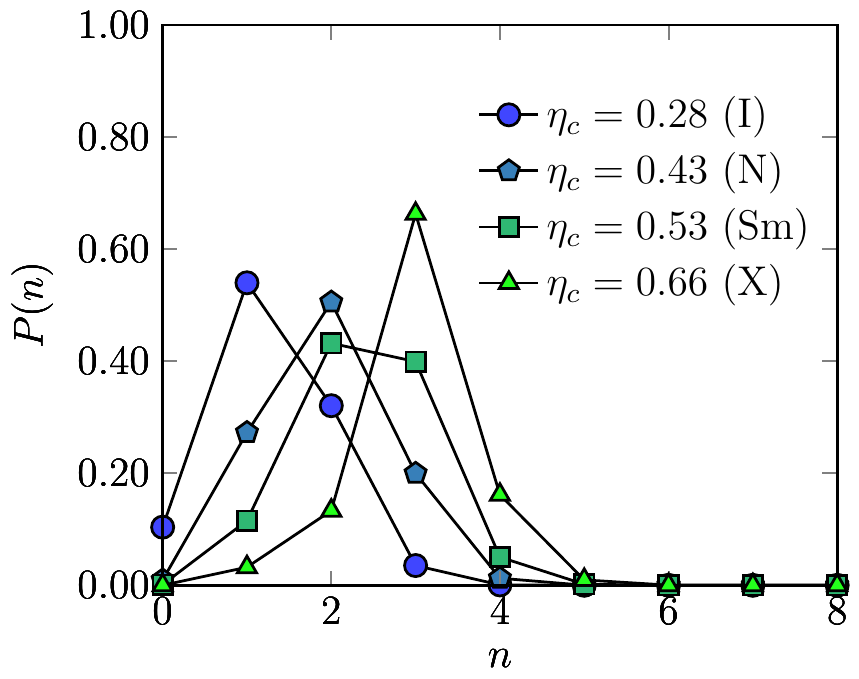}
\caption{\label{fig:Pn} The probability of $n$ overlapping depletion layers for a mixture of colloidal rods with length-to-diameter ratio $L/\sigma_c=5$ and non-adsorbing polymer of diameter $\sigma_p$ with size ratio $q=\sigma_p/\sigma_c=1$ at polymer fugacity $z_{p}=0$ and colloid packing fraction $\eta_{c}$ as labelled.}
\end{figure}

\begin{figure*}
\includegraphics[scale=0.20]{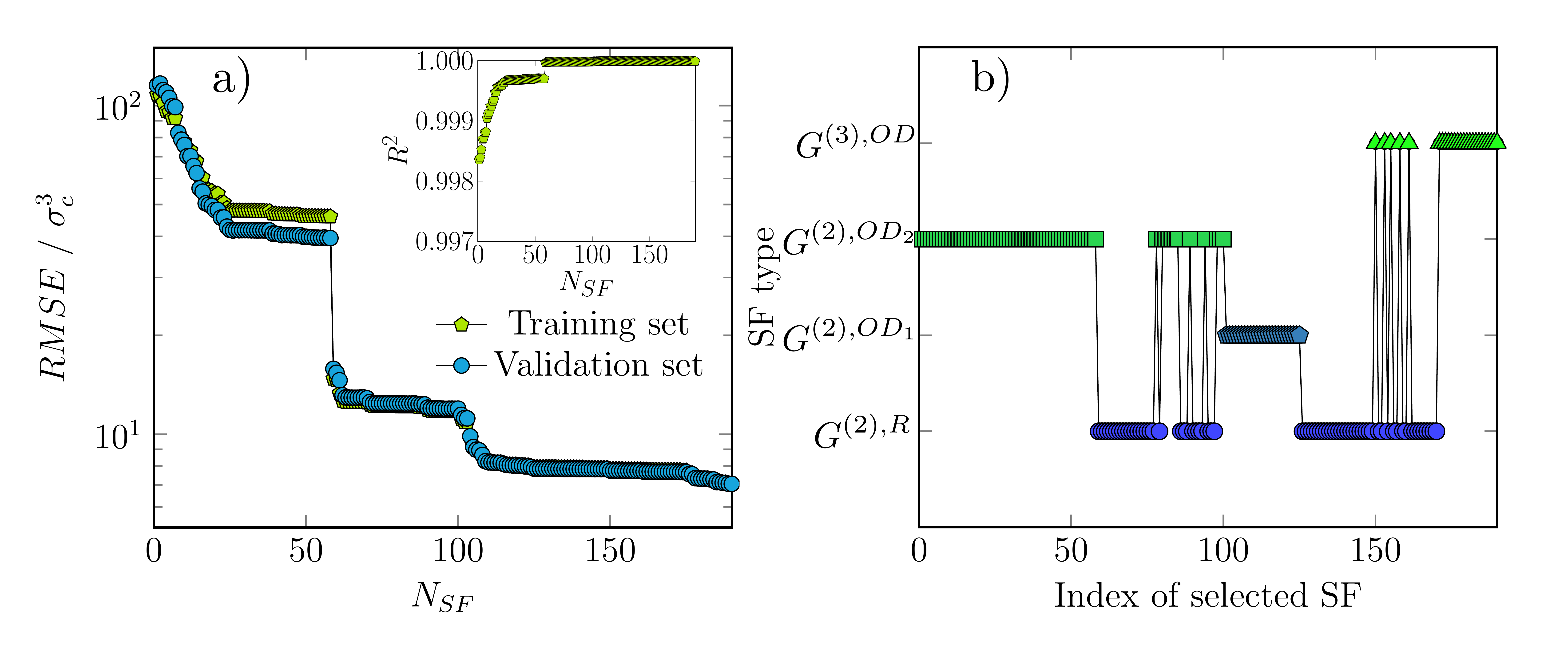}
\caption{\label{fig:fitmix} a) Root mean square error (RMSE) as a function of the number of SFs in the subset $N_{SF}$ for the effective one-component many-body Hamiltonian of a mixture of sterically-stabilized colloidal rods and non-adsorbing polymer coils. The RMSE values (in $\sigma_{c}^{3}$ units) are shown for the training and validation sets. The correlation coefficient $R^{2}$ as a function of $N_{SF}$ is shown in the inset. b) Type of  SF as a function of the index chosen in the feature selection method.}
\end{figure*}

To build the training data set, we perform MC simulations on $N_{c}=768$ colloids with $L/\sigma_{c}=5$, size ratio $q=1.0$,  polymer fugacity $z_{\text{p}}=0$, and colloid packing fraction $\eta_{\text{c}}=\pi  \sigma_{c}^{2} (2\sigma_{c}/3 + L) N_{c}/(4V) \in \left[ 0.13, 0.66\right]$ with a packing fraction spacing of $\delta \eta_{\text{c}}=0.005$.  From each simulation of $1\times 10^{7}$ MC cycles, consisting of $N_{c}$ attempts of rotating or translating particles, we collect 300 equilibrated, well-spaced configurations and measure $V_{f}$ using a numerical integration.~\cite{dijkstra2006effect,campos2021machine} The resulting data set contains a total of $27,900$ representative  particle configurations at different colloid densities, from which $80\%$ are used for training and $20\%$ for testing. Among the different thermodynamic states used for training, isotropic (I), nematic (N), smectic (Sm) and crystal (X) phases~\cite{bolhuis1997tracing}, which are characterized by different degrees of orientational and positional order, are considered. To quantify the importance of the many-body contributions to the effective potential in each of the stable phases, we calculate $P(n)$, the probability that we find  $n=n(\boldsymbol{r})$ overlapping depletion layers at spatial coordinate $\boldsymbol{r}$.~\cite{dijkstra2006effect,campos2021machine} In Fig.~\ref{fig:Pn} we show $P(n)$ for varying packing fraction $\eta_{c}$, including examples of I, N, Sm and X phases. In the high-density X phase at $\eta_{c}=0.66$, we find that even 4-body contributions to $V_f$ become non-negligible.



To fit $V_{f}$, we create a manageable pool of $M=365$ candidate SFs, setting the cut-off value at $r_c/\sigma_{c}=7.0$. Results of the fitting are reported in Fig.~\ref{fig:fitmix}. In particular, we show the RMSE  of the linear fits with the actual $V_{f}$ as a function of the number of selected SFs for both the training and the test sets. For $N_{SF}>104$, the correlation coefficient is $R^{2}\approx0.99$ and the normalized RMSE, defined as $\text{NRMSE} \equiv \text{RMSE}/|V_{f}^{\text{max}} -V_{f}^{\text{min}}|$, is on the order of $1\times 10^{-3}$, which clearly indicates the quality of the linear fit and ultimately, the ability of the ODSFs as fingerprints to encode information about the structure of the system of non-spherical particles. As described above, the feature selection method used here sequentially picks up the descriptors from the pool of candidate SFs based on the increase in linear correlation.  Fig.~\ref{fig:fitmix}b shows the type of descriptor that is sequentially selected from the pool of candidate SFs. We clearly observe that the $G^{(2),OD_{2}}(i; \alpha, R_s)$ descriptors describe best the variance of the data.

\begin{figure*}
\includegraphics[scale=0.20]{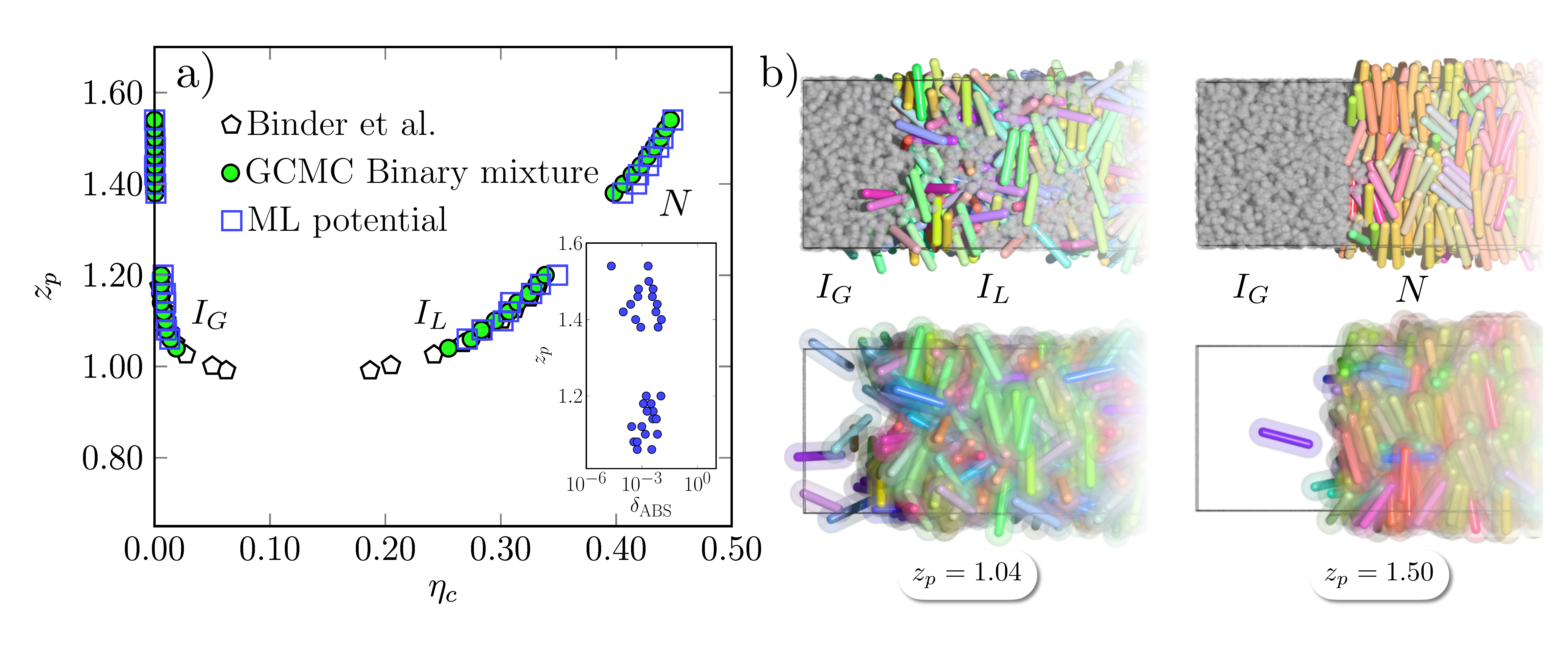}
\caption{\label{fig:coex} a) Phase diagram of a mixture of sterically-stabilized colloidal rods of length-to-diameter ratio $L/\sigma_{c}=5$ and non-adsorbing polymer coils of diameter $\sigma_{p}=1.0$ in the colloid packing fraction $\eta_c$-polymer fugacity $z_p$ representation. The size ratio of the mixture is $q=\sigma_p/\sigma_c=1$. Filled circles correspond to results obtained from grand-canonical Monte Carlo (GCMC)  simulations of the true binary mixture, whereas empty squares represent those obtained from direct coexistence Monte Carlo simulations using the fitted many-body ML potential. Empty pentagons represent previous results by Binder et al.~\cite{jungblut2007depletion} A broad coexistence region between an isotropic gas phase I$_{\text{G}}$ and an isotropic liquid phase I$_{\text{L}}$ is identified. At high polymer fugacity $z_p$, the I$_{\text{G}}$ phase is in equilibrium with a nematic N phase. The absolute errors ($\delta_{\text{ABS}}\equiv|\eta_{c,\text{ML}}-\eta_{c,\text{GCMC}}|$) of the coexisting colloid packing fraction predictions are shown in the inset as a function on the polymer fugacity $z_{p}$. b) Typical configurations of the fluid phase equilibria of the  colloid-polymer mixture as obtained from direct coexistence simulations using both  the full binary mixture (top) and the CG ML potential (bottom). The gray spheres correspond to the polymer coils and in the CG representation, the depletion layers of every spherocylinder are displayed. The colour of the rod-like particles is assigned according to their orientation in space.}
\end{figure*}

To validate the ML model and in order to assess its transferability, we calculate the (partial) phase diagram of the effective one-component system using canonical MC simulations. In particular, we focus on the binodals for the isotropic fluid phases.~\cite{savenko2006phase} To compute such binodals, we perform direct coexistence simulations of the effective one-component  system of $N_{c}=896$ colloids in elongated simulation boxes of volume $V$ with edges $L_{x}=L_{y} \geq 3L$ and $L_{z}=3L_{x}$  at varying polymer fugacity $z_{p}$. We fit the effective many-body Hamiltonian as evaluated in simulations on the full binary system with $N_{SF}=120$, which provides a NRMSE of $8\times 10^{-4}$. In order to confirm the accuracy of the fitted model, we also compute the coexistence curves from grand canonical Monte Carlo (GCMC) simulations of the full rod-sphere binary mixture, where $N_{c}$, $z_{p}$, $V$ and $T$ are kept constant. In Fig.~\ref{fig:coex} we report the phase diagrams in the ($\eta_{c},z_{p}$) plane as obtained from simulations of the two models. At sufficiently high polymer fugacity we find  a coexistence between a low-density isotropic gas  I$_{G}$ phase and a high-density isotropic liquid I$_{L}$  phase. We find a good correspondence between the two models and also agreement with previous results from GCMC simulations by Binder et al.~\cite{jungblut2007depletion}  For $z_{p}\geq 1.38$, a broad coexistence between the isotropic gas I$_{G}$ and nematic N phase is observed. Again we find  good agreement between the ML model and the "true" binary mixture. We highlight that since we do not include any two-phase systems with interfaces in our training set nor configurations at non-zero $z_{p}$, reproducing the coexisting curves with MC simulations of the ML model proves the transferability to finite $z_p$, but also that the ML potential captures the dependence on density that a realistic description must have to be able to calculate phase diagrams accurately.

\subsection{\label{sec:gel} Many-body interactions of core-shell microgel rods}

The third model we consider is a coarse-grained representation of core-shell microgel (CSM) rods  based on the models proposed in Refs.~\onlinecite{boattini2020modeling,pansu2017metallurgy}. These models capture the many-body interactions arising when elastic spheres are  strongly deformed  at high densities and high pressures.

\begin{figure*}
\includegraphics[scale=0.205]{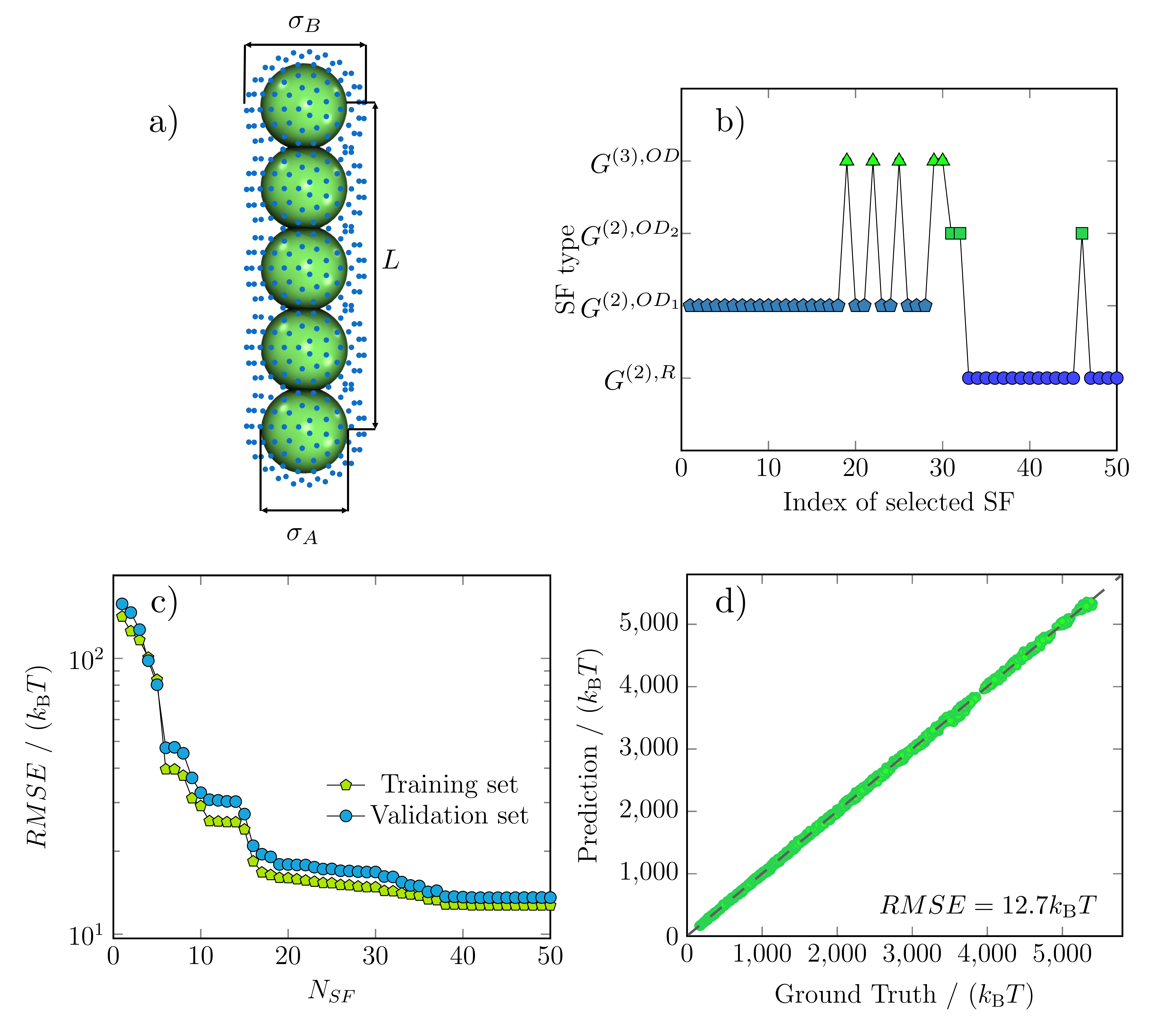}
\caption{\label{fig:fitmic} a) Schematic picture of a core-shell microgel (CSM) rod-like particle. The  core consisting of tangent hard spheres is depicted in green. The surface of the elastic microgel shell is shown in blue, where only the points on the surface are depicted for clarity. Below, the visual representation  of the deformation of the shell due to an overlap between two different rods (top view). b) Type of SF as a function of the index chosen in the feature selection method. c) The root mean square error, RMSE (in $k_{\text{B}}T$ units) as a function of the number $N_{SF}$, of SFs, where both the training and validation sets are plotted. The inset of c) shows the correlation coefficient $R^{2}$ as a function of $N_{SF}$.  d) Parity plot comparing the total elastic energy predictions of the ML model and the true model in configurations consisting of $N_{c}=768$ CSM rod-like particles.}
\end{figure*}

Here, we represent the core-shell microgel rod as a hard  spherocylindric core surrounded by a deformable shell. To represent the shape anisotropy of the particle, we describe the rod as an assembly of deformable spheres with rigid hard cores. The particle is constructed as follows: the rigid hard cores of the particles are represented in a `linear tangent' fashion,~\cite{vega2001liquid} where $n$ spherical hard cores of diameter $\sigma_A$ are linearly aligned and tangent to their neighbors as depicted in Fig.~\ref{fig:fitmic}a. Subsequently each rigid core is surrounded by a deformable shell. The  diameter of the rod is   $\sigma_B=\sigma_A+\lambda$, where $\lambda/2$ is the effective length of the deformable shell. 
In order to capture the elastic deformation due to an  interaction with another CSM particle, we discretize the surface of the deformable particle by a large number of  $N_p$ points, which are evenly distributed on the surface. When a microgel particle is deformed due to overlap with another particle, each point on the surface of this deformable shell will be pushed radially to the point of intersection of the two spheres. 
The elastic energy corresponding to the deformation of  particle $i$ is then  approximated by the weighted sum of the elastic energies associated to each point on its surface as~\cite{boattini2020modeling}
\begin{equation}
    \beta\phi_i=\frac{K}{2}\sum_{l=1}^{N_p}\frac{w_l}{\sigma_B^2}\left( \frac{\delta r_l}{\sigma_B} \right)^{2},
    \label{Eq:MGel}
\end{equation}
where $K$ is an elastic constant with the dimension of energy, $l$ runs over all the $N_p$ points on the surface of the deformable particle $i$, $w_l$ is the surface area of the associated Voronoi cell of point $l$ and $\delta r_l$ is the length of the radial deformation from the  surface of particle $i$ to the plane of intersection of the two spheres, as shown in Fig.~\ref{fig:fitmic}a. 
When particle $i$ overlaps with more than one particle at the same time, we evaluate for each  point $l$ on the surface of particle $i$  $\delta r^j_l$ for each interacting particle $j$, and use the maximum deformation to evaluate the elastic energy, i.e. the length of the radial deformation from the surface of particle $i$ to the intersection of all the spheres is  $\delta r_l=\max{\delta r^j_l}$.  

We consider a core-shell microgel rod with a rigid core consisting of  $n=5$ hard spheres. Following our notation of the hard-rod model, the length of the hard rod that forms the CSM particle is $L=4\sigma_A$ with semi-spherical caps at both ends with diameter $\sigma_A$. The elastic particle has a diameter of $\sigma_B=1.35\sigma_A$, corresponding to a $\lambda/2=0.175\sigma_A$. We discretize the surface of each sphere composing the rod with $N_p=200$ points, which we find  a good compromise between accuracy and efficiency. Finally, we set the elastic constant to $K=500 k_BT$.

 The computational cost of evaluating the elastic energy of  deformable spheres is very  high, as all the $N_p$ points on the surface of each sphere have to be taken into account when a Monte Carlo move is performed. The use of ML potentials to incorporate the many-body interactions of a 2D system of elastic spheres speeded up the energy evaluations considerably.~\cite{boattini2020modeling} The need of a ML technique for the present model is even more crucial as there is a dramatic increase of points $N_p=200\times5$ on the surface of each microgel rod to be evaluated. Hence, this model represents a perfect example where the orientation-dependent symmetry functions can be used to reduce the computational cost in simulating  these  systems. 

In order to apply the ODSFs to the CSM rod-like particles, we first build a data set based on the total elastic energy associated to configurations in different conditions. Due to the high computational cost of the model, we use the same equilibrated configurations as in Section~\ref{sec:mix} that correspond to $N_{c}=768$ rod-like particles. For each configuration, we calculate the elastic energy of each particle using Eq.~\ref{Eq:MGel} and subsequently the total elastic energy of the system. The training and validation configurations correspond to  various thermodynamic states ranging from  isotropic, nematic, smectic to crystalline phases.
We proceed as in the previous section, by generating a pool of ODSFs and selecting a subset of them that  capture best the variance of the target energy. Here, we use a cut-off of $r_{c}/\sigma_{A}=6.5$ and take $50$ ODSFs which give $R^{2}\approx 0.99$ and minimize the root mean square error on both training and validation sets, providing a $\text{NRMSE}$ of $2\times 10^{-3}$. In contrast with the model shown in the previous section, the variance of the data is best captured by the $G^{(2),OD_1}(i;\sigma_{||},\sigma_{\perp})$ descriptors as shown in Fig.~\ref{fig:fitmic}b. 
Note that with just a small number of ODSFs the data is already well described as the correlation coefficient approaches quickly to $R^2\approx0.99$ and the $\text{RMSE}$ is quickly minimized. Finally, we test the fitting by predicting the energies of the validation set. In Fig.~\ref{fig:fitmic}d we show the comparison of the predicted energy and the one calculated from Eq.~\ref{Eq:MGel}, we observe a perfect agreement between the two of them. We thus show that the ODSFs work well to capture the many-body elastic energy of CSM rod-like particles. This opens the door to explore the phase behavior and structural properties of complex systems such as CSM rods.

\subsection{\label{sec:CG} Two-body potential of mean force of ligand-stabilized nanorods}

\begin{figure*}
\includegraphics[scale=0.205]{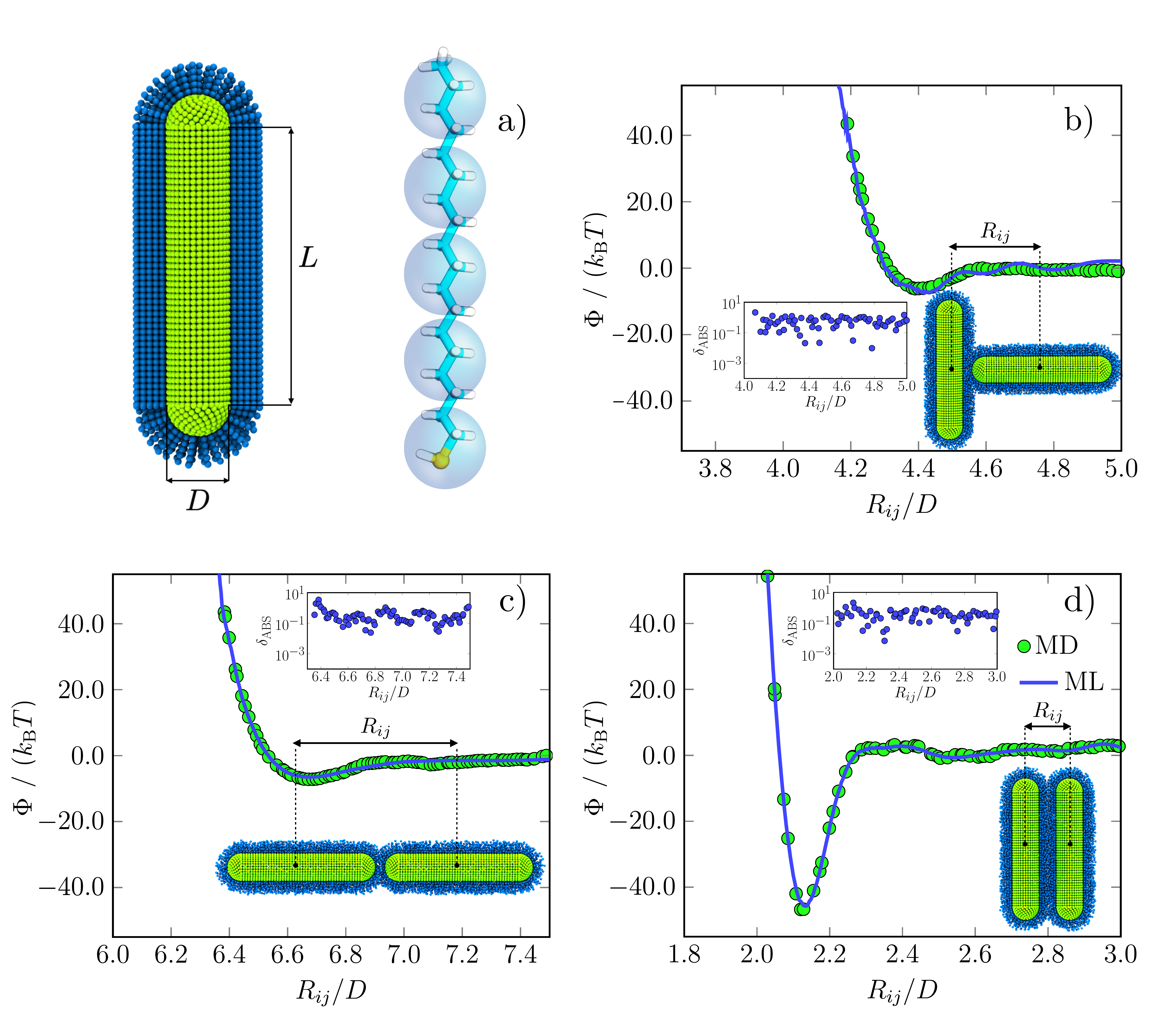}
\caption{\label{fig:PMF} a) Schematic representation of a ligand-stabilized nanorod of length $L$ and diameter $D$. Only a portion of the ligands is shown to better appreciate the morphology of the particles. The blue beads correspond to CG ligand atoms and the green beads represent the nanocrystal core atoms. In the CG representation, a single ligand (octadecanethiol) contains 5 CG beads which represent 18 carbon atoms (cyan) and a thiol group (yellow). b), c) and d) Potential of mean force $\Phi$ as a function of center of mass distance $R_{ij}$ for three different relative nanorod orientations, as illustrated in the snapshot of each plot. Green filled circles correspond to the values obtained from constraint MD simulations whereas blue lines correspond to the values predicted by the ML model. The absolute errors ($\delta_{\text{ABS}}\equiv|\Phi_{\text{ML}}-\Phi_{\text{MD}}|$) of the potential of mean force predictions are shown in the insets as a function center of mass distance $R_{ij}$.}
\end{figure*}

Finally we consider a chemistry-specific CG model of ligand-stabilized nanorods, for which we use the ODSFs to fit the effective two-body potential of mean force (PMF). In this model the ligands are represented as chains of 5 CG beads, approximately corresponding to alkyl ligands of 18 carbon atoms and a headgroup (e.g. thiol or amine), as shown in Fig.~\ref{fig:PMF}a. Nanocrystal cores (NCs) are modeled as rigid bodies composed of a cylinder of length $L=20$ nm and diameter $D=4.2$ nm, with semi-spherical caps at both ends. The orientation of a particle is thus determined by the central axis of its cylindrical core. The surface coverage, calculated as the number of ligand molecules per surface area is 5.5 nm$^{-2}$.\cite{monego2018colloidal} The CG ligands are covalently bonded to the nanocrystal cores and the interactions between the constituent CG beads are described by the MARTINI force field. For simplicity we consider only "C1"-type MARTINI beads. \cite{marrink2007martini} To account for the solvent implicitly, we use the approach by Fan and Gr\"unwald, \cite{fan2019orientational} where pair interactions between non-bonded beads are described through a modified Lennard-Jones (LJ) potential, which reads,

\begin{equation}
 \phi (r; s) =
    \begin{cases}
     \phi^{\textrm{LJ}}(r)+ (1-s)\epsilon~~ \text{for } r \leq 2^{1/6}\sigma\\
     s \phi^{\textrm{LJ}}(r) ~~~~~~~~~~~~~~~~~~\text{for }  2^{1/6}\sigma < r \leq r_{c},
    \end{cases}       
\end{equation}
where the quality of the solvent is controlled by the parameter $0\leq s \leq 1$ with $s=0$ corresponding to a good solvent and $s=1$ to a bad solvent or vacuum, and  
\begin{equation}
\phi^{\textrm{LJ}}(r) = 4 \epsilon \left[  \left(\frac{\sigma}{\epsilon} \right)^{12} - \left(\frac{\sigma}{\epsilon} \right)^{6} \right],
\end{equation}
is the standard Lennard-Jones (LJ) potential, $\sigma=0.47$ nm and $\epsilon=0.8365$ kcal mol$^{-1}$ are the length- and energy-scale parameters of the pair interaction, respectively, $r$ is the separation distance between pairs of coarse-grained sites, and $r_c=1.2$ nm is the cut-off radius of the interactions. We set the solvent parameter $s=0.3$. Within the chains, intramolecular interactions acting on the centres of bonded CG sites are described using a harmonic bond-stretching potential,
\begin{equation}
\phi^{\textrm{bond}}(b) = \frac{1}{2} K_{\text{b}} \left( b - b_{0} \right)^{2},
\end{equation}
\noindent with the bond force constant $K_{\text{b}}=149.3787$ kcal mol$^{-1}$nm$^{-2}$, and $b$ and $b_{0}=0.47$ nm the instantaneous and equilibrium bond distances, respectively. Similarly, the angle-bending between triplets of connected beads is modelled via a harmonic potential,
\begin{equation}
\phi^{\textrm{angle}}(\theta) = \frac{1}{2} K_{\theta} \left(  \cos \theta - \cos \theta_{0} \right)^{2},
\end{equation}
\noindent where $K_{\theta}=2.9876$ kcal mol$^{-1}$ denotes  the angle  force constant, and $\theta$ and $\theta_{0}=180^{\circ}$ the instantaneous and equilibrium angle-bending values, respectively. Interactions between NC cores are neglected as these forces, for small NCs, are typically much weaker than ligand-ligand interactions.\cite{schapotschnikow2009understanding, kister2018colloidal,widmer2014orientational} 
We perform molecular dynamics (MD) simulations on systems of 21132 CG beads using the software package LAMMPS.\cite{thompson2022lammps}  To avoid finite size effects, we employ a cubic simulation box of side length $10L$. We remove overlaps of ligand beads in the initial configurations  by energy minimization. Simulations are performed in the canonical ensemble ($NVT$), at temperature $T=300$ K, which is maintained constant with a Nos\'{e}-Hoover thermostat. The equations of motion are integrated with a time step of 20 fs, for up to 10$^7$ steps and statistics are collected over the last 10$^6$ steps.

To compute the effective pair interactions between the ligand-stabilized nanorods, we calculate the PMF for three different relative particle orientations, as shown in Fig.~\ref{fig:PMF}b, c and d, using constraint MD simulations.~\cite{ciccotti1989constrained} For each PMF calculation, we perform simulations with the nanorod cores frozen at 120 different distances $R_{ij}$. For each of these simulations, the mean force $F_{m}$ is calculated as the average force between the centres of mass of the two nanorods along the centre-of-mass distance vector $\textbf{R}_{ij}$~\cite{schapotschnikow2009understanding}
\begin{equation}
F_{m}(R_{ij}) = \frac{1}{2} \left< \left(\textbf{F}_i - \textbf{F}_j\right) \cdot \hat{\textbf{R}}_{ij}\right>_{NVT; R_{ij},\hat{\boldsymbol{u}_{i}}, \hat{\boldsymbol{u}_{j}}}
\end{equation}
where $\textbf{F}_i$ and $\textbf{F}_j$ are the total forces acting on the centre of mass of each core and $\hat{\textbf{R}}_{ij}$ is the unit vector connecting the two rods along the reaction coordinate $R_{ij}$. Angular brackets denote ensemble averages in the canonical ensemble with the constraint  separation $R_{ij}$ and orientations $\hat{\boldsymbol{u}_{i}}$ and $\hat{\boldsymbol{u}_{j}}$ of the NCs. The PMF $\Phi(R_{ij})$ is then computed using 
\begin{equation}
\Phi(R_{ij}) = \int^{\infty}_{R_{ij}} F_{m}(R_{ij}) dR'_{ij}.
\end{equation}
In Fig.~\ref{fig:PMF}b, c and d, the computed PMF curves clearly show that the effective interactions between the nanorods are repulsive at short distances and attractive at larger distances. In particular, we find that when the particles are parallel aligned, the strength of the effective attractive interaction is the strongest, with a deep local minimum of $-46.7~k_{\text{B}}T$ at $R_{ij}/D=2.1$. The value of the well depth for the side-by-side configuration is around 7 times larger than that of the other two configurations.

To employ the ML approach explained above, we combine the data points of the three measured PMFs into one training data set that includes particle-particle distances with corresponding orientations and free energies associated to them. We perform the fits using two-body ODSFs with a cut-off of $r_{c}/D=8.0$ and we restrict the number of descriptors to $N_{SF}=100$. In the Supplementary Material, we show the RMSE as a function of the number of SFs and the type of SF as a function of the index chosen in the feature selection scheme. The ML models constructed with the two ODSFs introduced in Section~\ref{sec:model}, present a correlation coefficient $R^{2} \approx 0.99$ and a $\text{NRMSE}$ of $7.8\times10^{-3}$. To further test the accuracy of the ML model, we evaluate $\Phi$ using a single interaction site representation for the particles. In particular, for each of the three relative orientations, we generate up to 500 configurations at fixed distance $R_{ij}$ and evaluate the PMF using the fitted model. Considering that the generated particle configurations are all different from those included in the original training data set, the agreement with the values obtained from the MD simulations (see Fig.~\ref{fig:PMF}b, c and d) highlights the ability of the ML model to accurately interpolate between structures and smoothly recover (predict) $\Phi$. This coarse-graining strategy, which is based on a mapping that projects fine-grained configurations of the complex nanorods onto a lower-dimensional representation (a single site model), can be used to efficiently explore the phase behavior and structural properties by long simulations of large system sizes. The quality of the fitted CG potential will obviously reflect that of the underlying fine-grained model. Thus, its predictive power and accuracy would strongly depend on the quality of the training data set (e.g. number of training samples and the numerical precision of the measured PMF). Furthermore, as discussed above, for a given data set, the quality of the fitted CG potential can be controlled in our approach by the number of descriptors $N_{SF}$.

\section{\label{sec:conclusions}Conclusions}
In summary, we have proposed new orientation-dependent particle-centered descriptors that effectively map a static configuration of anisotropic rod-like particles into a suitable representation which can be employed to construct a machine learning model to regress a structure-property relation. To demonstrate the ability of the functions in describing orientation and alignment effects, we have used simple linear regression to construct an effective single-component Hamiltonian for hard colloidal rods and non-adsorbing polymer by formally integrating out the polymers and fitting the grand potential (free volume) that incorporates many-body effects. The resulting ML potential was used in direct coexistence simulations to calculate the phase diagram of the mixture. We found good agreement with results obtained from simulations of the true binary mixture. Additionally, an accurate and computationally-efficient many-body interaction potential of anisotropic core-shell microgel particles has been fitted on the basis of the proposed descriptors. The same approach has also been used to represent the effective orientation-dependent two-body potential of mean force of a chemistry-specific nearly-atomistic model of ligand-stabilized nanorods.

The methodology presented here can be seen as a bottom-up coarse-graining strategy valuable for speeding up simulations of complex anisotropic particles. The speedup achieved with the ML potentials depends on the details of the precise underlying fine-grained model and the type and number of descriptors that are used for the construction of the ML potentials. Roughly, we find that the ML potentials are 20 times slower for the case of Gay-Berne ellipsoids, three orders of magnitude faster for the colloid-polymer mixtures, and three times faster for the microgel particles (see Supplementary Material). It would be interesting to extend this approach to construct CG models with anisotropic building blocks (superatoms, beads) that are able to capture the large-scale properties of specific molecular systems, in the same spirit as the ellipsoid-based models developed for semiflexible polymers.~\cite{lee2012ellipsoid}  This strategy may also allow one to move from idealized to predictive models of mesogenic molecules, relevant in the field of liquid crystals. Furthermore, the proposed functions could be exploited in a ML approach to address how orientational structure and shape anisotropy determine the dynamics of elongated particles. Finally, we note that the functions discussed here are perfectly suited for describing the orientational and translational degrees of freedom of anisotropic uniaxial particles with elongated or oblate shape. Therefore, it would be of interest to develop new descriptors suited for mapping the configurations of non-spherical particles with more complex symmetry (e.g. biaxial particles).

\section*{SUPPLEMENTARY MATERIAL}

See supplementary material for a complete list of parameters used to construct the initial pool of candidate SFs for building the CG potentials for the different models, learning curves and a brief discussion on the computational efficiency of the ML potentials. Additionally, simple example FORTRAN functions to evaluate the SFs are also provided.

\begin{acknowledgments}
G.C.-V. acknowledges funding from The Netherlands  Organization  for  Scientific  Research  (NWO) for the ENW PPS Fund 2018 – Technology Area Soft Advanced Materials ENPPS.TA.018.002. G.G. acknowledges funding from the Netherlands Center for Multiscale Catalytic Energy Conversion (MCEC), a NWO Gravitation program funded by the Ministry of Education, Culture and Science of the government of the Netherlands. M.D. and S.M.-A. receive funding from the European Research Council (ERC) under the European Union's Horizon 2020 research and innovation programme (Grant agreement No. ERC-2019-ADG 884902 SoftML).

\end{acknowledgments}

\section*{Conflict of interest}
The authors have no conflicts to disclose.

\section*{Data Availability}
The data that support the findings of this study are available from the corresponding author upon reasonable request.

\bibliography{aipsamp}

\end{document}